\documentclass[journal]{IEEEtran}

\usepackage[utf8]{inputenc} %

\usepackage{parskip} %
\usepackage{graphicx} %
\usepackage[caption=false]{subfig}

\usepackage{amsmath,amssymb,amsfonts} %
\usepackage{bm, bbm} %
\usepackage{amsthm} %
\usepackage{algorithm} %
\usepackage[noend]{algpseudocode} %

\DeclareMathOperator*{\argmax}{arg\,max}

\usepackage[nopatch=footnote]{microtype} %
\usepackage{comment} 
\usepackage[commandnameprefix=always]{changes}
\usepackage{soul} %
\usepackage{orcidlink} %
\usepackage{setspace}
\usepackage{adjustbox}

\usepackage{array} %
\usepackage{longtable}
\usepackage{multirow}
\usepackage{booktabs}

\usepackage{arydshln} %

\usepackage{cite} %
\usepackage[capitalise]{cleveref}
\crefname{appsec}{Appendix}{Appendices}
\crefname{equation}{}{}
\usepackage{jabbrv}

\usepackage{xcolor}
\usepackage{color} %
\definecolor{darkred}{rgb}{0.6,0.0,0.0}
\definecolor{darkgreen}{rgb}{0,0.50,0}
\definecolor{lightblue}{rgb}{0.0,0.42,0.91}
\definecolor{orange}{rgb}{0.99,0.48,0.13}
\definecolor{grass}{rgb}{0.18,0.80,0.18}
\definecolor{pink}{rgb}{0.97,0.15,0.45}
\definecolor{mypagecolor}{RGB}{255,255,255}
\pagecolor{mypagecolor}
\definecolor{mytextcolor}{RGB}{0,0,0}
\AtBeginDocument{\color{mytextcolor}}

\usepackage{hyperref}
\hypersetup{
colorlinks = true,
linkcolor = black,
urlcolor = blue,
citecolor = blue
}

\usepackage{bookmark}
\usepackage[acronym, toc]{glossaries}

\usepackage[acronym, toc]{glossaries}
\newacronym{RB}{RB}{resource block} 
\newacronym{IoT}{IoT}{Internet of things}
\newacronym{MTC}{MTC}{machine-type communication}
\newacronym{mMTC}{mMTC}{massive machine-type communication}
\newacronym{IIoT}{IIoT}{industrial Internet of Things}
\newacronym{HTC}{HTC}{Human-type communication}
\newacronym{MA}{MA}{multiple access}
\newacronym{PA}{PA}{priority access}
\newacronym{RACH}{RACH}{random access channel}
\newacronym{PRACH}{PRACH}{pysical random access channel}

\newacronym{PG}{PG}{power grid}
\newacronym[\glslongpluralkey={smart grids}, \glsshortpluralkey={SGs}]{SG}{SG}{smart grid} %
\newacronym{DOE}{DOE}{Department of Energy}
\newacronym{AMI}{AMI}{advanced metering infrastructure}
\newacronym{SM}{SM}{smart meter}
\newacronym{HAN}{HAN}{home area network}
\newacronym{DR}{DR}{demand response}
\newacronym{EV}{EV}{electric vehicle}
\newacronym{WASA}{WASA}{wide-area situational awareness}
\newacronym{DER}{DER}{distributed energy resources}
\newacronym{DGM}{DGM}{distributed energy management}
\newacronym{NIST}{NIST}{National Institute of Standards and Technology}
\newacronym{FERC}{FERC}{Federal Energy Regulatory Commission}

\newacronym{UE}{UE}{user equipment}
\newacronym{CB-MA}{CB-MA}{contention-based multiple access}
\newacronym{CF-MA}{CF-MA}{conflict-free multiple access}
\newacronym{FDMA}{FDMA}{frequency division multiple access}
\newacronym{TDMA}{TDMA}{time division multiple access}
\newacronym{CDMA}{CDMA}{code division multiple access}
\newacronym{SDMA}{OMA}{spatial division multiple access}
\newacronym{LDS-CDMA}{LDS-CDMA}{low density signature code division multiple access}
\newacronym{PDMA}{PDMA}{pattern division multiple access}
\newacronym{MUSA}{MUSA}{multi-user shared access}
\newacronym{SCMA}{SCMA}{sparce code multiple access}
\newacronym{IDMA}{IDMA}{interleave division multiple access}
\newacronym{SPMA}{SPMA}{statistical based multiple access}
\newacronym{DS-CDMA}{DS-CDMA}{direct sequence code division multiple access}
\newacronym{FH-CDMA}{FH-CDMA}{frequency hopping code division multiple access}

\newacronym{OMA}{OMA}{orthogonal multiple access}
\newacronym{NOMA}{NOMA}{non-orthogonal multiple access}
\newacronym{CD-NOMA}{CD-NOMA}{code domain non-orthogonal multiple access}
\newacronym{PD-NOMA}{PD-NOMA}{power domain non-orthogonal multiple access}
\newacronym{MIMO}{MIMO}{multiple input multiple output}
\newacronym{QAM}{QAM}{quadrature amplitude modulation}
\newacronym{FEC}{FEC}{forward error correction}
\newacronym{MAC}{MAC}{medium access control}
\newacronym{SIC}{SIC}{successive interference cancellation}
\newacronym{BS}{BS}{base station}
\newacronym{LTE}{LTE}{long-term evolution}
\newacronym{LTE-A}{LTE-A}{long-term evolution advanced}

\newacronym{SE}{SE}{spectral efficiency}
\newacronym{CSI}{CSI}{channel state information}
\newacronym{SUD}{SUD}{single-user detection}
\newacronym{MUD}{MUD}{multi-user detection}
\newacronym{MAP}{MAP}{maximum \textit{a posteriori}}
\newacronym{ML}{ML}{maximum likelihood}
\newacronym{RCML}{RCML}{reduced complexity maximum likelihood}
\newacronym{i.i.d.}{i.i.d.}{independent and identically distributed}

\newacronym{FDD}{FDD}{frequency division duplexing}
\newacronym{TDD}{TDD}{time division duplexing}
\newacronym{PSD}{PSD}{power spectral density}
\newacronym{MAI}{MAI}{multiple access interference}
\newacronym{QoS}{QoS}{quality of service}
\newacronym{TTNT}{TTNT}{Tactical Targeting Network Technology}
\newacronym{COS}{COS}{channel occupancy statistic}
\newacronym{PPP}{PPP}{Poisson point process}
\newacronym{3GPP}{3GPP}{3rd Generation Partnership Project}

\newacronym{H-UE}{H-UE}{high-priority user equipment}
\newacronym{L-UE}{L-UE}{low-priority user equipment}
\newacronym{ACB}{ACB}{access class barring}
\newacronym{D-ACB}{D-ACB}{dynamic access class barring}

\newacronym{QPAC}{QPAC}{queue-aware priority access classification}
\newacronym{PRADA}{PRADA}{prioritized random access with dynamic access barring}

\newacronym{RHS}{RHS}{right hand side}
\newacronym{LHS}{LHS}{left hand side}
\newacronym{MAE}{MAE}{mean absolute error}
\newacronym{SQP}{SQP}{sequential quadratic programming}

\newacronym{RL}{RL}{reinforcement learning}
\newacronym{MDP}{MDP}{Markov decision process}
\newacronym{MAB}{MAB}{multi-armed bandit}
\newacronym{UCB}{UCB}{upper confidence bound}
\newacronym{AS}{AS}{action space}
\newacronym{CAS}{CAS}{compact action space}
\newacronym{e-greedy}{$\epsilon$-greedy}{epsilon-greedy}
\newacronym{DEG}{DEG}{decaying epsilon-greedy}
\newacronym{DQN}{DQN}{deep Q-network}
\newacronym[\glsshortpluralkey={algs.}]{alg}{alg.}{algorithm}
\newacronym{CE}{CE}{cross entropy}
\newacronym{PPO}{PPO}{proximal policy optimization}

\begin{document}
\title{Access Probability Optimization in RACH: \\ A Multi-Armed Bandits Approach}

\author{
	Ahmed~O.~Elmeligy~\orcidlink{0009-0001-5969-1323},
	Ioannis~Psaromiligkos~\orcidlink{0000-0002-1643-5143},~\IEEEmembership{Member,~IEEE,}
	and~Au~Minh~\orcidlink{0000-0003-2745-0055}
	\thanks{A. Elmeligy and I. Psaromilgkos are with the Department of Electrical and Computer Engineering, McGill University, Montreal, QC, Canada, e-mails: ahmed.elmeligy@mail.mcgill.ca.,  ioannis.psaromiligkos@mcgill.ca.}
	\thanks{Au Minh is with Hydro-Quebéc Research Institute (IREQ), Varennes, QC, Canada, e-mail: au.minh2@hydroquebec.com.}
}

\markboth{}%
{}
\maketitle
\begin{abstract}
	The use of cellular networks for massive machine-type communications (mMTC) is an appealing solution due to the availability of the existing infrastructure. 
	However, the massive number of user equipments (UEs) poses a significant challenge to the cellular network's random access channel (RACH) regarding congestion and overloading.
	To mitigate this problem, we first present a novel approach to model a two-priority RACH, which allows us to define access patterns that describe the random access behavior of UEs as observed by the base station (BS).
	A non-uniform preamble selection scheme is proposed, offering increased flexibility in resource allocation for different UE priority classes.
	Then, we formulate an allocation model that finds the optimal access probabilities to maximize the success rate of high-priority UEs while constraining low-priority UEs.
	Finally, we develop a reinforcement learning approach to solving the optimization problem using multi-armed bandits, which provides a near-optimal but scalable solution and does not require the BS to know the number of UEs in the network.
\end{abstract}
\IEEEpeerreviewmaketitle

\setlength{\textfloatsep}{0pt}
\setlength{\floatsep}{0pt}
\setlength{\intextsep}{1pt}

\section{Introduction}
\label{sec: intro}
\IEEEPARstart{T}{he} \gls{RACH} is a crucial component in cellular networks, enabling \glspl{UE} to request access for communication and establish a connection to the \gls{BS}~\cite{wei2014modeling}
The RACH is created through a periodic allocation of time and frequency resources, known as \glspl{RACH} slot~\cite{laya2013random}.
The periodicity of the \gls{RACH} slot is determined by the physical \gls{RACH} configuration index, which is broadcasted by the \gls{BS}.
During a \gls{RACH} slot, \glspl{UE} can access the network by transmitting a randomly selected preamble to the \gls{BS} from a finite set of preambles.
A preamble is a sequence of symbols uniquely identifying a \gls{UE} to the \gls{BS}; it can be thought of as a \gls{UE}'s digital signature.

If a single \gls{UE} chooses a unique preamble, the \gls{BS} can successfully decode it.
However, challenges arise in the context of \gls{mMTC} applications, where a large number of \glspl{UE} accessing the network through the \gls{RACH} can lead to congestion and overloading issues \cite{condoluci2015toward, wiriaatmadja2014hybrid}.
Futhermore, congestion can be exacerbated by recurring \glspl{UE} that periodically synchronize to communicate simultaneously with application servers~\cite{3gpp-37.869}.
The \gls{RACH} congestion problem arises when multiple \glspl{UE} select the same preamble during a slot, resulting in collisions \cite{arouk2016accurate}.
In essence, this situation can be likened to a scenario where a finite set of \glspl{RB}, represented by the preambles, are randomly accessed by a pool of \glspl{UE}.
The collisions pose a significant challenge in effectively managing the access requests from a massive number of \glspl{UE}~\cite{swain2022novel, kurnia2018random}.

The challenge intensifies due to the implementation of \gls{PA}, which consolidates multiple \gls{mMTC} applications to be served by a single cellular network.
This strategic approach aims for cost efficiency by leveraging existing infrastructure and eliminating the necessity for deploying multiple networks.
However, the amalgamation of diverse \gls{mMTC} applications seeking \gls{PA} introduces complexities in the \gls{RACH} mechanism, necessitating innovative approaches to mitigate congestion and enhance the overall efficiency of the access procedure.

\subsection{Related Work}
\label{subsec: related work}
There have been several studies on the \gls{PA} problem in the literature.
\Gls{SPMA} is a \gls{PA} scheme first utilized by the U.S. Military \gls{TTNT}~\cite{clark_statistical_2010}. 
In \gls{SPMA}, each \gls{UE} first estimates how busy the channel is by calculating the \gls{COS}. 
Several methods have been developed to calculate the \gls{COS}~\cite{clark_statistical_2010, yang2018modeling}; for example, to find the \gls{COS}, each \gls{UE} counts the number of transmissions heard from other \glspl{UE} in the network using a weighted sum over a predefined statistical sliding window~\cite{liu_performance_2017}. 
Then, the \gls{COS} is compared to a threshold that depends on the \gls{UE}'s priority. 
If the \gls{COS} is below the threshold, the \gls{UE} can access the channel. 
Otherwise, the \gls{UE} is denied access. 
Once channel access is granted, \gls{SPMA} uses a combination of frequency and time hopping in conjunction with turbo codes to split the packet into multiple bursts and transmit them over the channel~\cite{zhang_modeling_2022}. 

On the other hand, the work in~\cite{althumali2022priority} dynamically adjusts the number of available \glspl{RB} for each priority class based on the network load before each \gls{RACH} slot to reduce access delay.
Another \gls{PA} scheme is \Gls{ACB} proposed by the \gls{3GPP} to mitigate access congestion in \gls{LTE} random access~\cite{duan2016d}. 
In \gls{ACB}, each \gls{UE} is assigned a backoff time or barring rate depending on the predetermined priority level of the \gls{UE}. 
A challenge with \gls{ACB} is that the barring rate for each \gls{UE} is fixed; as a result, if a \gls{L-UE} is denied access multiple times to the channel, its buffer might overflow, resulting in dropped packets. 
A solution to the above problem is the \gls{QPAC}-based \gls{ACB}, which dynamically adjusts the barring rate based on the packet queue size of each \gls{UE}~\cite{chowdhury_queue-aware_2022}. 
In \gls{QPAC}, a higher priority is assigned to \glspl{UE} with longer packet queues in their respective buffers. 
\textcolor{black}{The authors in \cite{liu2020online} dynamically allocate non-overlapping \glspl{RB} to each priority class by first estimating the number of \glspl{UE} in each class using a Bayesian approach. 
This estimation is based on the number of empty \glspl{RB}, leaving out potentially useful information that could improve the estimator's accuracy.
Based on the estimation, the \glspl{RB} are then distributed to achieve a success rate that aligns with the corresponding priority class.}

Moving beyond conventional approaches, several studies have explored the application of \gls{RL} techniques to optimize resource allocation in cellular networks~\cite{zehra2022utilizing, pacheco2019deep, nguyen2022reinforcement, yang2022reinforcement}.
\gls{RL}, a subfield of machine learning, involves training agents via trial and error to make decisions in an environment to achieve specific goals.
Through continuous interaction and adaptation, \gls{RL} enables autonomous decision-making, making it well-suited for addressing the complex and evolving nature of resource allocation challenges in cellular networks, where the network is subject to constant changes in \gls{UE} behavior, traffic patterns, network conditions, and overall demand for resources.
For instance, in the context of cellular networks, Q-learning has been employed to maximize spectrum utilization efficiency~\cite{liu2022reinforcement}, while in~\cite{shinkafi2021priority}, learning automata in combination with Q-learning is used to minimize \gls{UE} collisions.
Furthermore, the authors in~\cite{zhang2022ppo} leverage a proximal policy optimization \gls{RL} algorithm to dynamically adjust the number of available \glspl{RB} and find \gls{ACB} parameters that maximize the number of successful \glspl{UE}.
Deep \gls{RL} approaches, as seen in~\cite{gedikli2022deep}, aim to reduce collisions and improve the success probability of \glspl{UE} by dynamically distributing \glspl{RB} across different priority classes instead of using a fixed allocation such as in~\cite{cheng2011prioritized, lin_prada_2014, kalalas2016handling}; this is possible due to the ability of deep \gls{RL} to leverage deep neural networks and learn complex and non-linear hierarchical relationships between the network's state and the optimal action to take.
Additionally, distributed \gls{RL} is applied in~\cite{bai2021multiagent} to each \gls{UE}'s \gls{RB} selection policy to minimize the access delay.
\textcolor{black}{The authors in \cite{orim2023random} mitigate congestion by aggregating data from multiple UEs before transmission. Q-learning is employed to determine the optimal amount of data to aggregate before initiating the random access process, effectively reducing the number of UEs contending for access.}

These \gls{RL} techniques showcase promising avenues for enhancing the efficiency and performance of cellular networks in dynamic and complex environments.

\subsection{Novelty and Contributions}
\label{subsec: novelty and contributions}
All the above \gls{PA} schemes assume that the \gls{UE} randomly selects a \gls{RB} from a uniform distribution.
To our knowledge, there is a gap in the existing literature regarding a model that considers non-uniform preamble selection, which offers increased flexibility in resource allocation across different \gls{UE} priority classes. 
The flexibility of non-uniform preamble selection allows each priority class to achieve the desired \gls{QoS} metrics by adjusting the access probabilities.
It is noteworthy that our investigation, as detailed in \cref{sec: results}, demonstrates that the optimal solution, i.e., the access probabilities that maximize the success rate of high-priority \glspl{UE} while constraining low-priority \glspl{UE}, involves a non-uniform preamble selection approach. 
This highlights the practical importance of considering other access probability distributions beyond the conventional uniform distribution, as it can significantly enhance the performance of cellular networks.

Specifically, the main contributions of this paper are as follows:

\begin{itemize}
    \item We introduce a novel dimension to the field by accommodating non-uniform preamble selection within the same \gls{RACH} slot. This approach generalizes the existing literature that assumes uniform access probabilities and allows us to find the optimal access probabilities for each priority class.
	\item We present a two-priority \gls{RACH} model, which allows us to define access patterns that describe the random access behavior of \glspl{UE} as observed by the BS in terms of the locations of \gls{H-UE} successes, \gls{L-UE} successes, collisions, and empty \glspl{RB} in \gls{RACH} slots.
	\item We propose an algorithm that finds the optimal access probabilities to maximize the success rate of \glspl{H-UE} while constraining \glspl{L-UE}.
	\item We develop a \gls{MAB} \gls{RL} approach to solving the access probabilities optimization problem, which provides a near-optimal but scalable solution and does not require the \gls{BS} to know the number of \glspl{UE} in the network. Our \gls{RL} method employs a systematic approach that initially discretizes the \gls{AS}, then reduces its size. Subsequently, the method incorporates reward scaling to refine the learning process further~\cite{fujimoto2021minimalist}. We utilize cross-entropy for the \gls{MAB} updates, which is particularly advantageous for large-scale scenarios, as it does not necessitate trying out every single action, thereby significantly improving computational efficiency.
	\item \textcolor{black}{We present an alternative \gls{CAS} that incorporates a lookup table with the optimal access probabilities of varying network, allowing us to compress the \gls{AS} further and as a byproduct estimate the network load. However, the tradeoff involves introducing additional hyperparameters, which would need to be tuned based on a rough estimate of the network load.}
\end{itemize}

This paper is structured as follows:
\cref{sec: system model} presents the two-priority \gls{RACH} model.
The optimal access pattern allocation problem is formulated in \cref{sec: RACH optimization}.
\cref{sec: multi-armed bandits} presents the \gls{MAB} formulation to solve the optimization problem.
Finally, \cref{sec: results,sec: conclusion} present the numerical results and conclude the paper, respectively.

\section{System Model and Problem Statement}
\label{sec: system model}
We consider a system model with $n^h$ \glspl{H-UE} and $n^l$ \glspl{L-UE} assuming that $n^h$ and $n^l$ are fixed but  unknown to the \gls{BS}.
Time is discretized into slots, indexed by $t$, and each slot is divided into $M$ \glspl{RB}.
At each slot $t$, we assume all $n = n^h + n^l$ \glspl{UE} have a packet to transmit and that each \gls{UE} randomly chooses an \gls{RB} to transmit its packet.
By denoting $p^h_i$ and $p^l_i$ the respective probabilities that a \gls{H-UE} chooses \gls{RB} $i$ and an \gls{L-UE} chooses \gls{RB} $i$, let $\bm{p}^h = [p^h_1, \dots, p^h_M]$ and  $\bm{p}^l = [p^l_1, \dots, p^l_M]$ be vectors collecting the \gls{RB} access probabilities for a \gls{H-UE}, and an \gls{L-UE}, respectively.
Clearly, we have $\sum_{i=1}^M p^h_i = 1$ and $\sum_{i=1}^M p^l_i = 1$. 
We assume the access probability vectors $\bm{p}^h$ and $\bm{p}^l$ are known at the \gls{BS}, which is responsible for generating and broadcasting these probabilities to the \glspl{UE}.

At each of the $M$ \glspl{RB}, one of the following events may occur: 1)Event $H_i$: A single \gls{H-UE} selects \gls{RB} $i$. 2) Event $L_i$: A single \gls{L-UE} selects \gls{RB} $i$. 3) Event $\Phi_i$: No \gls{UE} selects \gls{RB} $i$, i.e., the \gls{RB} is unoccupied or empty. 4) Event $X_i$: More than one \gls{UE} selects \gls{RB} $i$, i.e., a collision occurs.

\begin{figure}
	\centering
	\begin{adjustbox}{max width=0.4\textwidth}
	\includegraphics{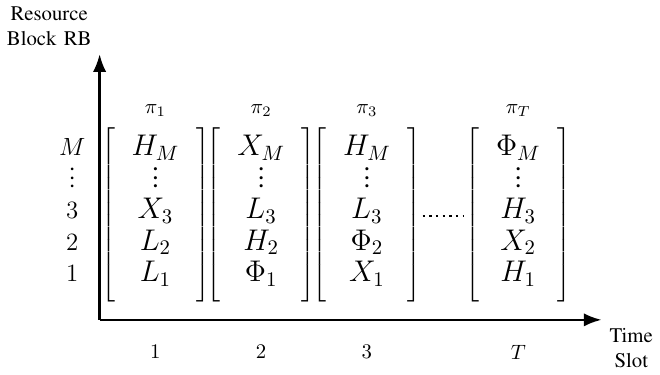}
	\end{adjustbox}
	\centering \caption{Example of a pattern sequence over $T$ time slots.}
	\label{fig: patterns}
\end{figure}

Each event can be detected by the \gls{BS}, however, in the case of a collision, the \gls{BS} does not know how many \glspl{UE} are involved~\cite{wu2012fast, galinina2013stabilizing, wang2015optimal, zhang2022ppo}. 
Thus, at time slot $t$, the \gls{BS} observes an access pattern that is the sequence of events over the $M$ \glspl{RB}.
An example of such patterns observed over $T$ time slots is shown in \cref{fig: patterns}.
It is assumed that $\bm{\pi}_1, \bm{\pi}_2, \ldots, \bm{\pi}_T$ are \gls{i.i.d.}.

The problem we consider in this paper is to find the optimal access probability vectors $\bm{p}^h$ and $\bm{p}^l$ that maximize the \gls{H-UE} \gls{RACH} throughput while guaranteeing that the \gls{L-UE} \gls{RACH} throughput is above a certain threshold.
In the next section, we first define the \gls{H-UE} \gls{RACH} and \gls{L-UE} \gls{RACH} throughput metrics, and then formally describe the optimization problem.

\section{Access Probability Optimization Formulation}
\label{sec: RACH optimization}
We will start by developing the performance metrics, using which we will then formulate the optimization problem.
For this section, we assume that the \gls{BS} knows the network load, $n^h$ and $n^l$.
This assumption is relaxed in the next section when we develop a \gls{MAB} approach to solve the optimization problem.

\subsection{Throughput Metrics}
\label{subsec: metric development}
The \gls{RACH} throughput is defined as the average number of UEs per RACH slot that successfully avoid transmission collisions~\cite{althumali2022priority}.
Since two priority classes was defined in this paper, we consider two throughputs $\mu_h$ and $\mu_l$ as the average number of successful \glspl{H-UE} and \glspl{L-UE} per \gls{RACH} slot respectively
In such a condition, by denoting $H_{\bm{\pi}}$ and $L_{\bm{\pi}}$ the number of successful \glspl{H-UE} and \glspl{L-UE} for a given access pattern $\bm{\pi}$, the expected number of successful \glspl{H-UE} and \glspl{L-UE} per \gls{RACH} slot, are given by
\begin{equation}
	\label{eq: high priority throughput}
	\mu_h = \sum_{\bm{\pi} \in \Pi} H_{\bm{\pi}} P(\bm{\pi})
\end{equation}
\begin{equation}
	\label{eq: low priority throughput}
	\mu_l = \sum_{\bm{\pi} \in \Pi} L_{\bm{\pi}} P(\bm{\pi})
\end{equation}
where $\Pi$ is the set of all possible access patterns.
However, finding $\Pi$ is non-trivial, especially for large networks.
Hence, we develop a recursive backtracking algorithm, the details of which are omitted for brevity.

\textcolor{black}{The calculation of $P(\bm{\pi})$ represents the probability of observing the access pattern $\bm{\pi}$.
Let sets $\mathcal{H}$, $\mathcal{L}$, $\mathbf{\mathit{\Phi}}$, and $\mathcal{X}$ represent the indices of \glspl{RB} with events $H_i$, $L_i$, $\Phi_i$, and $X_i$, respectively.
Thus, $\bm{\pi}$ is the intersection of events $H_i$ for $i \in \mathcal{H}$, $L_i$ for $i \in \mathcal{L}$, $\Phi_i$ for $i \in \mathbf{\mathit{\Phi}}$, and $X_i$ for $i \in \mathcal{X}$. 
The probability $P(\bm{\pi})$ is then found from this intersection.
\begin{equation}
    \label{eq: conditional pattern probability}
    P(\bm{\pi}) = 
    P\left(\overline{H}\right)
    P\left(\overline{L} \middle| \overline{H}\right)
    P\left(\overline{\Phi} \middle| \overline{H}, \overline{L}\right)
    P\left(\overline{X} \middle| \overline{H}, \overline{L}, \overline{\Phi}\right)
\end{equation}
where $\overline{H}=\bigcap_{i\in{\mathcal H}}H_i$, $\overline{L}=\bigcap_{i\in{\mathcal L}}L_i$, $\overline{\Phi}=\bigcap_{i\in\mathbf{\mathit{\Phi}}}\Phi_i$, and $\overline{X}=\bigcap_{i\in{\mathcal X}}X_i$.
More detailed derivation of the factors in \cref{eq: conditional pattern probability} can be found in our previous work~\cite{elmeligy2024load}.}

\subsection{Problem Formulation for Optimal Access Pattern Allocation}
\label{subsec: problem formulation}
This section formulates the optimization problem based on the previously developed two-priority \gls{RACH} model.
We focus on maximizing the expected \gls{H-UE} \gls{RACH} throughput, $\mu_h$,
under the constraint that the \gls{L-UE} \gls{RACH} throughput, $\mu_l$, is greater than or equal to a specified threshold, $\gamma$.
\footnote{Otherwise, the \gls{L-UE} \gls{RACH} throughput will be zero, and the network will be reduced to a single-priority network.}
The optimization problem is formulated as a constrained non-linear continuous optimization problem, as shown in \crefrange{eq: optimization problem}{const: 4}.
Constraints \cref{const: 1,const: 2,const: 3,const: 4} ensure that $\bm{p}^h$ and $\bm{p^l}$ are probability vectors.

\begin{subequations}
    \begin{align}
        \label{eq: optimization problem}
		\max_{\bm{p}^h, \bm{p}^l} \quad & \mu_h \\
		\text{s.t.} \quad & \mu_l \geq \gamma \\
		& \sum_{i=1}^{M} p^h_i = 1 \label{const: 1}\\
		& \sum_{i=1}^{M} p^l_i = 1 \label{const: 2}\\
		& p^h_i \geq 0 \quad \forall i \in \{1, 2, \dots, M\} \label{const: 3}\\
		& p^l_i \geq 0 \quad \forall i \in \{1, 2, \dots, M\}   \label{const: 4}
    \end{align}
\end{subequations}

The optimization problem can be solved using a gradient-based optimization solver implemented such as the \gls{SQP} algorithm, which has been proven to be a robust algorithm that efficiently solves non-linear optimization problems; Schittkowski~\cite{schittkowski1986nlpql} showed that \gls{SQP} surpasses other optimization methods in efficiency, precision, and robustness.
As we will see in \cref{subsec: RACH optimization results}, the above optimization formulation does not scale well in terms of time complexity.
Thus, we propose a novel approach to solve the optimization problem using \gls{MAB}.

\section{Multi-Armed Bandits Problem Formulation}
\label{sec: multi-armed bandits}
The \gls{MAB} problem is a sequential decision-making problem where an agent repeatedly chooses from a set of actions, known as arms, in a stochastic environment.
Each arm has an unknown reward distribution, and the agent's goal is to maximize the reward over time.
This is done by utilizing a Q-value for each action, which serves as an estimate of the expected reward associated with choosing that action.
Q-values enable the agent to make informed decisions based on past experiences.
\gls{MAB} is a simplified version of the \gls{RL} problem since the agent does not have to consider the state of the environment when selecting an action.

The agent must balance exploration and exploitation to obtain a good estimate of the Q-value associated with each action. We denote an action by $A$, and the set of all possible actions as \gls{AS}.
Each time the agent performs $A$ it observes a reward $r$. 
Then, the Q-value for $A$, $Q(A)$, is updated as follows:
\begin{equation}
	\label{eq: Q update}
	Q(A) = Q(A) + \frac{1}{v[A]} \left( r - Q(A) \right)
\end{equation}
where $v[A]$ is the number of times $A$ has been performed.

\subsection{Action Space and Reward}
For the problem at hand, we define $A$ as a pair of access probability vectors $\bm{p}^h$ and $\bm{p}^l$.
The elements in $\mathbf{p}^h$ and $\mathbf{p}^l$ are continuous, which means that $A$ is uncountable.
Therefore, the \gls{MAB} problem formulation is not directly applicable to our problem as it assumes that the \gls{AS} is discrete and finite.
This assumption is made for several reasons. 
First, many classic \gls{MAB} algorithms, such as \gls{e-greedy}, \gls{DEG}, \gls{UCB}, and softmax, are designed for discrete \glspl{AS}.
Second, discrete \glspl{AS} lower the computational complexity  and improve the memory efficiency of algorithms since look-up tables can store and manage information about the values or rewards associated with each action.
This is particularly important when dealing with large-scale problems.
In \Cref{subsec: AS discretization} we will present two discrete and finite \glspl{AS} tailored to our problem.

Our goal is to find the optimal access probability vectors $\bm{p}^h$ and $\bm{p}^l$ that maximize the \gls{H-UE} \gls{RACH} throughput while maintaining a minimum \gls{L-UE} \gls{RACH} throughput.
Thus, we define the reward, $r$, as follows:
\begin{equation} 
	\label{eq: reward}
	r = \begin{cases}
		\mu_{h, T} & \text{if } \mu_{l, T} \geq \gamma \\
		\rho \mu_{h, T} & \text{otherwise}
	\end{cases}
\end{equation}
where $\rho \geq 0$ is a penalty factor, $T$ is the number of \gls{RACH} slots, and $\mu_{h, T}$ and $\mu_{l, T}$, are the numerical expectations of the \gls{RACH} throughputs, respectively, given by
\begin{align}
	\label{eq: high priority throughput T}
	\mu_{h, T} &= \frac{1}{T} \sum_{t=1}^{T} \sum_{m=1}^M \mathbbm{1}(RB_i = h) \\
	\label{eq: low priority throughput T}
	\mu_{l, T} &= \frac{1}{T} \sum_{t=1}^{T} \sum_{m=1}^M \mathbbm{1}(RB_i = l)
\end{align}
In the above, $RB_i$ represents the $i$-th \gls{RB} in the \gls{RACH} slot while, $\mathbbm{1}(RB_i = h)$ and $\mathbbm{1}(RB_i = l)$ are indicator functions that return one if the \gls{RB} $i$ is occupied by a single \gls{H-UE} or \gls{L-UE}, respectively, and zero otherwise.
In~\cref{eq: reward}, the reward is $\mu_{h, T}$ if the \gls{L-UE} \gls{RACH} throughput is above the threshold $\gamma$; otherwise, the reward is $\mu_{l, T}$ multiplied by the penalty factor $\rho$ to penalize actions that result in $\mu_{l, T}< \gamma$.

In our definition of the reward, we did not use the throughputs $\mu_h$ and $\mu_l$, defined by \cref{eq: high priority throughput,eq: low priority throughput}, respectively, for two reasons: First, unlike $\mu_h$ and $\mu_l$, $\mu_{h, T}$ and $\mu_{l, T}$ do not require the knowledge of $n^h$ and $n^l$.
Second, numerical expectations are more suitable for \gls{MAB} as they significantly reduce the time and computational resources required to simulate the \gls{RACH} procedure.

\subsection{Action Space Formulations}
\label{subsec: AS discretization}
We will consider two distinct \gls{AS} sets: A discretized \gls{AS}, and a \gls{CAS}, which are explained below.

\subsubsection{Discretized Action Space}
\label{subsubsec: discrete AS}
The \gls{AS} can be discretized by a step size, $d$, which determines the granularity of the \gls{AS}: A smaller step size results in a higher resolution \gls{AS} that captures more possible actions but also results in a larger \gls{AS} size and vice versa.
Although the above formulation is suitable for \gls{MAB} problems, it becomes computationally infeasible for large values of $M$ and small values of $d$.
Therefore, we need to reduce the \gls{AS} size to make the problem computationally tractable.

We observe that circular shifts of the elements in $\mathbf{p}^h$ and $\mathbf{p}^l$ result in the same access probabilities, which leads to redundant actions that do not need to be considered.
In such case, the \gls{AS} size can be reduced by simply eliminating redundant actions such as $[[0.3, 0.3, 0.4], [0.2, 0.3, 0.5]]$ and $[[0.4, 0.3, 0.3], [0.5, 0.3, 0.2]]$. for example.

\cref{tab:discretization-step-size-and-action-space-size} compares the \gls{AS} size before and after removing the redundant actions, for different values of $M$ and $d$.
We see that the \gls{AS} size is significantly reduced, especially for large values of $M$ or small values of $d$ when the circular shift check is applied.
For example, for $M=5$ and $d=0.2$, the \gls{AS} size is reduced from 15876 to 3176, a reduction of 80\%.
The implementation of the \gls{AS} discretization and reduction is presented in Appendix~\ref{app: AS discretization algorithms}.
We note that it can be done offline, and the resulting \gls{AS} can be stored for use during the online \gls{RL} process.

\textcolor{black}{The \gls{AS} formulation described earlier is well-suited for \gls{MAB} algorithms that do not require exhaustive exploration of the entire \gls{AS} to converge, such as the \gls{CE} algorithm adopted in this paper and outlined in \cref{subsec: formulation as a multi-armed bandit problem}.}
However, even with redundancy elimination, traditional \gls{MAB} algorithms can become computationally prohibitive as the \gls{AS} size increases exponentially with an increase in $M$ and polynomially with a reduction in $d$.
Even for a large step size of 0.5, the \gls{AS} size is still significant, and is computationally impractical to explore with traditional \gls{MAB} algorithms; for example, \gls{UCB} requires sweeping the entire \gls{AS} to converge~\cite{auer2002finite}, while other methods attempt to estimate the value of each action~\cite{burtini2015survey}, which is impractical with a large \gls{AS}.

\begin{table}[t]
	\centering
	\footnotesize
	\centering \caption{\gls{AS} and reduced \gls{AS} size, after redundancy elimination, for different $M$ and $d$.}
	\label{tab:discretization-step-size-and-action-space-size}
	\begin{tabular}{@{}llll@{}}
		\toprule
		$M$ & $d$ & \textbf{\gls{AS} size} & \textbf{Reduced \gls{AS} size}\\ \midrule
		\multirow{3}{*}{2} & 0.5 & 9 & 5 \\ \cdashline{2-4}[0.3pt/2pt]
		& 0.2 & 36 & 18 \\ \cdashline{2-4}[0.3pt/2pt]
		& 0.1 & 121 & 61 \\ \hdashline[0.6pt/1pt]
		\multirow{3}{*}{3} & 0.5 & 36 & 12 \\ \cdashline{2-4}[0.3pt/2pt]
		& 0.2 & 441 & 147 \\ \cdashline{2-4}[0.3pt/2pt]
		& 0.1 & 3844 & 1452 \\ \hdashline[0.6pt/1pt]
		\multirow{2}{*}{4} & 0.5 & 100 & 26 \\ \cdashline{2-4}[0.3pt/2pt]
		& 0.2 & 3136 & 784 \\ \hdashline[0.6pt/1pt]
		\multirow{2}{*}{5} & 0.5 & 225 & 45 \\ \cdashline{2-4}[0.3pt/2pt]
		& 0.2 & 15876 & 3176 \\ \bottomrule
	\end{tabular}
\end{table}

\subsubsection{Compact Action Space}
\label{subsubsec: compact AS}
We now propose an alternative approach to construct a reduced-size \gls{AS}.
In this approach, the \gls{AS} includes the optimal access probabilities for each pair $(n^h, n^l)$ where $n^h$ ranges from $0$ to $N^h_s$ and $n^l$ from $0$ to $n^l = N^l_s$, with $N^h_s$ and $N^l_s$ denoting the maximum number of \gls{H-UE} and \gls{L-UE} to consider, respectively.
The optimal access probabilities are calculated offline inadvance by solving the optimization problem in \cref{subsec: problem formulation}, and saved in a look-up table for future use by the \gls{MAB} optimization process.

An interesting byproduct of this approach is an estimate of the network load, which can be easily obtained as the values of $n^h$ and $n^l$ that correspond to $\bm{p}^h$ and $\bm{p^l}$ determined by the MAB method.
The accuracy of this network load estimate is investigated in \cref{subsec: MAB estimation}.

The size of the \gls{CAS} is $N^h_s \times N^l_s$, which is significantly smaller than the \gls{AS} size obtained by the above discretization and reduction methods, provided that $N^h_s$ and $N^l_s$ are chosen carefully: Large enough to capture the true values of $n^h$ and $n^l$ but small enough to ensure that the \gls{AS} size is manageable.
\textcolor{black}{However, selecting appropriate values for $N^h_s$ and $N^l_s$ is challenging and requires a rough prior estimation of the network load. Although the approach ultimately provides an accurate network load estimate, this prerequisite remains a critical consideration.}

\textcolor{black}{Selecting between the discretized \gls{AS}, and a \gls{CAS} requires careful consideration of the specific application, network size and characteristics. 
For the discretized \gls{AS}, the level of granularity (i.e. the discretization step) is a critical factor, determining how close the actions align with the constraints threshold, $\gamma$.
Applications requiring high precision or strict adherence to constraints will benefit from finer granularity, although this increases the computational overhead.}

\textcolor{black}{On the other hand, adopting a \gls{CAS} relies on a rough prior estimate of the network load, as these influence the choice of hyperparameters like $N^h_s$ and $N^l_s$.
While this prior estimate is a key consideration, it is not directly addressed in this work.
The \gls{CAS} approach is particularly well-suited for scenarios where precomputing optimal access probabilities for manageable subsets of \gls{H-UE} and \gls{L-UE} is feasible, as it balances computational efficiency and network performance. 
Ultimately, the choice of the \gls{AS} hinges on finding the right balance between computational complexity, precision, and practical application requirements.
Results for both the discrete action space and \gls{CAS} approaches are presented in \cref{sec: results}, highlighting their respective tradeoffs and applicability.}

\subsection{Reward Scaling}
\label{subsec: reward scaling}
Most \gls{RL} algorithms show a performance increase when adopting a bounded reward~\cite{li2023internally, strouse2021learning}.
This approach simplifies the hyperparameter tuning process as the reward scale is fixed~\cite{sullivan2024reward}.
Moreover, limiting the reward signal facilitates learning across different tasks using a single algorithm and hyperparameter configuration, as in the original \gls{DQN} paper, where the reward signal is clipped to the range $[-1, 1]$~\cite{mnih2013playing}.
Additionally, scaling the reward can improve the stability of the learning process, as the reward signal is less likely to explode or vanish~\cite{mnih2015human,naeem2020generative}.
Constraining the range of rewards prevents overly large or small rewards from dominating the learning process and disproportionately influencing the policy.

In this paper, $n^h$ and $n^l$ can be very large and may experience frequent and unpredictable fluctuations depending on the specific system or environment. 
Scaling the reward can help ensure the system's scalability and adaptability under diverse operating conditions while maintaining similar hyperparameter configurations.
The most common approach for scaling the reward is to normalize it to the range $[0, 1]$.
However, this approach requires knowledge of the maximum reward which can be challenging to obtain. 
For example, in the two-priority \gls{RACH} model, as will be shown in \cref{subsec: RACH optimization results}, the maximum value of  $\mu_{h, T}$ can be easily determined only when $n^h \leq M-1$ by setting $\mathbf{p}^h$ and $\mathbf{p}^l$ to \cref{eq: p scaling,eq: x scaling}, respectively.
\begin {equation}
	\label{eq: p scaling}
	p^h_i = \begin{cases}
		\frac{1}{M-1} & \text{if } 1 \leq i \leq M-1 \\
		0 & \text{if } i = M
	\end{cases}
\end{equation}
\begin {equation}
	\label{eq: x scaling}
	p^l_i  = \begin{cases}
		0 & \text{if } 1 \leq i \leq M-1 \\
		1 & \text{if } i = M
	\end{cases}
\end{equation}
When $n^h > M-1$, the maximum reward is not straightforward to determine.
So, instead of normalizing $\mu_{h, T}$, we propose to scale the reward by dividing $\mu_{h, T}$ by the value of $\mu_h$ obtained when $\mathbf{p}^h$ and $\mathbf{p}^l$ are set to \cref{eq: p scaling,eq: x scaling}, respectively.

To maximize $\mu_h$, when there is no constraint on $\mu_l$, we need to set $\mathbf{p}^h$ to a uniform distribution across $M-1$ \glspl{RB} and allocate all \glspl{L-UE} to the remaining \gls{RB}.
This strategy achieves the maximum reward when $n^h \leq M-1$.
In this case, the scaling factor aligns with the maximum achievable reward, resulting in a maximum scaled value of 1.
However, for $n^h > M-1$, using the same strategy for $\mathbf{p}^h$ and $\mathbf{p}^l$ causes $\mu_h$ to drop due to increased contention among \glspl{UE} across the $M-1$ RBs.
To mitigate this drop, a non-uniform allocation strategy is required, which adjusts $\mathbf{p}^h$ and $\mathbf{p}^l$ to better handle the higher load, as will be demonstrated in \cref{subsec: RACH optimization results}.
Nevertheless, since the scaling factor remains based on a less efficient strategy, i.e. uniform allocation, it does not fully reflect the improvements made by the new non-uniform allocation strategy.
Therefore, the $\mu_h$ appears slightly greater than 1 when scaled.

The advantages of scaling the reward as described above are similar to those of normalizing the reward to the $[0, 1]$ range, with the added benefit of avoiding the task of determining the maximum reward when $n^h > M-1$.
This contrasts the priority-based \gls{RACH} model, developed in \cref{sec: system model}, where $\mu_h$ and $\mu_l$, are bounded to the range $[0, n^h]$ and $[0, n^l]$, respectively.
We are now ready to utilize the priority-based \gls{RACH} model to formulate the \gls{MAB} problem used in optimizing the access probability vectors.

\subsection{Formulation as a Multi-Armed Bandit Problem}
\label{subsec: formulation as a multi-armed bandit problem}
The priority-based \gls{RACH} model can be incorporated in the \gls{MAB} problem formulation as shown in \cref{alg: MAB formulation}.
A \gls{CE} approach based on the work in~\cite{wang2017cemab} is used to solve the \gls{MAB} problem, as \gls{CE} is more suitable for large-scale problems than other \gls{MAB} algorithms~\cite{rubinstein2004cross}.
The algorithm takes several inputs: $M$ and $T$ are the \gls{RACH} simulation parameters, where $T$ is the number of time slots as defined in \cref{sec: system model}.
The \gls{MAB} hyperparameters include $\gamma$, $d$, runs, $B_s$, $s$, and $\alpha$, where $B_s$ is the batch size, $s$ is the used to determine the size of the 'elite' batch, $B_e$, and $\alpha$ is the learning rate.

\begin{algorithm}[t]
	\centering \caption{\gls{MAB} formulation}
	\label{alg: MAB formulation}
	\footnotesize
	\begin{algorithmic}[1]
		\Function{MAB}{$M$, $T$, $\gamma$, $d$, runs, $B_s$, $s$, $\alpha$}
		\State $\mathcal{AS} \gets \texttt{GenerateActionSpace}(d)$ \label{alg: MAB AS generation}
		\State $\mathbf{Q} \gets \mathbf{0}_{1 \times |\mathcal{AS}|}\text{,} \mathbf{V} \gets \mathbf{0}_{1 \times |\mathcal{AS}|} \text{, \& } \mathbf{p}^{\mathcal{AS}} \gets (\mathbf{\frac{1}{|\mathcal{AS}|}})_{1 \times |\mathcal{AS}|}$ \label{alg: MAB initialization}
		\State $N_B \gets \text{int}(\frac{\text{runs}}{B_s}) \text{ \& } B_e \gets \text{int}(s B_s)$
			\For {\_ in $\text{range}(N_B)$} \label{alg: MAB main loop start}
				\State $\mathbf{D} \gets \text{empty list}$ \label{alg: MAB D initialization}
				\For {\_ in $\text{range}(B_s)$}
					\State $\mathbf{A} \gets \texttt{SelectAction}(\mathcal{AS}, \mathbf{p}^{\mathcal{AS}}$) \label{alg: MAB action selection}
					\State $\mathbf{V}[\mathbf{A}] \gets \mathbf{V}[\mathbf{A}] + 1$
					\State $\mu_{h, T}, \mu_{l, T}\gets \text{from \cref{eq: high priority throughput T,eq: low priority throughput T}}$ \label{alg: MAB RACH simulation}
					\If {$\mu_{l, T}< \gamma$} \label{alg: MAB constraint start}
						\State $r \gets \rho \times \texttt{scale}(\mu_{h, T})$ \label{alg: MAB penalty}
					\Else
						\State $r \gets \texttt{scale}(\mu_{h, T})$ \label{alg: MAB no penalty}
					\EndIf \label{alg: MAB constraint end}
					\State $\mathbf{Q}[\mathbf{A}] \gets \mathbf{Q}[\mathbf{A}] + \frac{1}{\mathbf{V}[\mathbf{A}]} (r - \mathbf{Q}[\mathbf{A}])$ \label{alg: MAB Q update}
					\State $\mathbf{D}.\text{append}(idx_{\mathbf{A}},  \mathbf{Q}[\mathbf{A}])$ \label{alg: MAB D}
					\EndFor
					\State $\tilde{\mathbf{p}}^{\mathcal{AS}} \gets \texttt{CeUpdate}(|\mathcal{AS}|, B_e, \mathbf{D})$ \label{alg: MAB ce update}
				\State $\mathbf{p}^{\mathcal{AS}} \gets (1 - \alpha) \mathbf{p}^{\mathcal{AS}} + \alpha \tilde{\mathbf{p}}^{\mathcal{AS}}$ \label{alg: MAB p update}
				\EndFor \label{alg: MAB main loop end}
			\State $idx_{\mathbf{A}} \gets \argmax_{\mathbf{A}} \mathbf{Q}[\mathbf{A}]$ \label{alg: MAB best A}
			\State \Return $\mathcal{AS}[idx_{\mathbf{A}}]$ \label{alg: MAB return}
		\EndFunction
	\end{algorithmic}
\end{algorithm}

The \gls{AS} is initially generated by one of the methods outlined in \cref{subsubsec: discrete AS,subsubsec: compact AS} (\cref{alg: MAB AS generation}).
Vectors $\mathbf{Q}$ and $\mathbf{V}$, both of length $|\mathcal{AS}|$, are used to store $Q[A]$ and $v[A]$ for each action, respectively. 
The probabilities of selecting each action are stored in $\mathbf{p}^{\mathcal{AS}}$, which is also a vector of length $|\mathcal{AS}|$.
Both $\mathbf{Q}$ and $\mathbf{V}$ are initialized to zero, while $\mathbf{p}^{\mathcal{AS}}$ is initialized to a uniform distribution (line~\ref{alg: MAB initialization}).
Subsequently, the algorithm determines the number of batches, $N_B$, and the size of the 'elite' batch, $B_e$ (line~\ref{alg: MAB initialization}).

Next, the algorithm enters the main loop, which runs for $N_B$ iterations (lines~\ref{alg: MAB main loop start} to \ref{alg: MAB main loop end}).
In each iteration, a vector $\mathbf{D}$, which contains the index of the action and the corresponding $\mathbf{Q}$, and is used to update $\mathbf{p}^{\mathcal{AS}}$, is first initialized to an empty list (line~\ref{alg: MAB D initialization}).
Then, a subloop runs for $B_s$ iterations, where an action, $\mathbf{A}$, is drawn from the \gls{AS} using the probability distribution $\mathbf{p}^{\mathcal{AS}}$ (line~\ref{alg: MAB action selection}).
The number of times each action is visited is updated in $\mathbf{V}$, and the \gls{RACH} simulation is run to obtain $\mu_{h, T}$ and $\mu_{l, T}$.

Afterwards, the \gls{RACH} is observed for $T$ time slots, and $\mu_{h, T}$ and $\mu_{l, T}$ are calculated using \cref{eq: high priority throughput T,eq: low priority throughput T} (line~\ref{alg: MAB RACH simulation}).
The details of the \gls{RACH} simulation are provided in Appendix~\ref{app: RACH simulation}.

The reward is then obtained using \cref{eq: reward} (lines~\ref{alg: MAB penalty} to \ref{alg: MAB no penalty}) and scaled using the scaling approach described in \cref{subsec: reward scaling}.
The vector $\mathbf{Q}$ is updated next using the reward signal (line~\ref{alg: MAB Q update}).
Following this, $\mathbf{D}$ is updated with the index of the action, $idx_{\mathbf{A}}$, and the corresponding $\mathbf{Q}$ (line~\ref{alg: MAB D}).

After the subloop is completed, the algorithm updates $\mathbf{p}^{\mathcal{AS}}$ by first generating a new probability distribution, $\tilde{\mathbf{p}}^{\mathcal{AS}}$, using the \gls{CE} update (line~\ref{alg: MAB ce update}), which is described in the next paragraph.
Then, $\mathbf{p}^{\mathcal{AS}}$ is smoothed using the new probability distribution $\tilde{\mathbf{p}}^{\mathcal{AS}}$ to avoid drastic changes in the probability distribution (line~\ref{alg: MAB p update}).
Finally, the algorithm returns the action with the highest $\mathbf{Q}$ value (lines~\ref{alg: MAB best A} and \ref{alg: MAB return}).

The distribution $\mathbf{p}^{\mathcal{AS}}$ is updated at the end of each batch using \gls{CE} in an asynchronous manner, which is described in \cref{alg: ceupdate}.
The algorithm only updates the probability distribution of the 'elite' batch, $B_e$, a subset of the batch with the highest $\mathbf{Q}$ values, and the rest of the batch is ignored~\cite{wang2017cemab}.
Vector $\mathbf{D}$ is first sorted, in descending order of $\mathbf{Q}$ (line~\ref{alg: ceupdate sort}).
Then, the algorithm generates $\tilde{\mathbf{p}}^{\mathcal{AS}}$, by using an indicator function, $\mathbf{I}$, which counts the number of times each action is selected in the 'elite' batch (line~\ref{alg: indicator}).

\begin{algorithm}[t]
	\centering \caption{\Gls{CE} update}
	\label{alg: ceupdate}
    \footnotesize
	\begin{algorithmic}[1]
		\Function{CeUpdate}{$|\mathcal{AS}|$, $B_e$, $\mathbf{D}$}
			\State $\tilde{\mathbf{p}}^{\mathcal{AS}} \gets \mathbf{0}_{1 \times |\mathcal{AS}|}$
			\For {$idx_{\mathbf{A}}$ in $\text{range}(|\mathcal{AS}|)$}
				\State $\mathbf{Q}_{sort}\text{, } \mathbf{idx}_{\mathbf{Q}_{sort}} \gets \texttt{Sort}(\mathbf{D})$ \label{alg: ceupdate sort}
				\State $\tilde{\mathbf{p}}^{\mathcal{AS}}[idx_{\mathbf{A}}] \gets \frac{1}{B_e} \sum_{i=1}^{B_e} \mathbf{I}(\mathbf{idx}_{\mathbf{Q}_{sort}}[i] = idx_{\mathbf{A}})$ \label{alg: indicator}
			\EndFor
			\State \Return $\tilde{\mathbf{p}}^{\mathcal{AS}}$
			
		\EndFunction
	\end{algorithmic}
\end{algorithm}

\section{Results}
\label{sec: results}
In this section we examine the performance of the proposed optimization methods.
\textcolor{black}{We begin by evaluating the uniform \gls{RB} allocation for \glspl{H-UE} and \glspl{L-UE}, as well as \gls{ACB} scheme, both serving as benchmarks. The results for uniform allocation and \gls{ACB} are presented in \cref{subsec: uniform access probabilities} and \cref{subsec: ACB}, respectively.}
Afterwards, we present the results for the optimal access probability method in \cref{subsec: problem formulation}, followed by the results for the \gls{MAB}-based method in \cref{sec: multi-armed bandits}.
Finally, we analyze the accuracy of estimating the network load using the \gls{CAS} as discussed in \cref{subsubsec: compact AS}.
Unless stated otherwise, the number of \glspl{H-UE} and \glspl{L-UE} are set to 4 and 5, respectively, and the number of \glspl{RB} is varied from 3 to 6.
We note that the hardware used for the simulations is an Intel Core i7-1065G7 CPU (1.3 GHz) with 16 GB of RAM.
We note that the hardware used for the simulations is an Intel Core i7-1065G7 CPU (1.3 GHz) with 16 GB of RAM.

\subsection{Uniform Access Probabilities}
\label{subsec: uniform access probabilities}
The results for the uniform \gls{RB} allocation are shown in \cref{tab: uniform optimization results}.
\begin{table}
	\footnotesize
	\centering
	\caption{Uniform access probabilities results with $n^h=4$, $n^l=5$.}
	\label{tab: uniform optimization results}
	\begin{tabular}{@{}llllll@{}}
		\toprule
		~ & \multicolumn{2}{l}{\textbf{Output Variables}} & \multicolumn{2}{l}{\textbf{Metrics}}\\ \cmidrule(lr){2-3} \cmidrule(lr){4-6}
		$M$ & $\mathbf{p}^h$ & $\mathbf{p}^l$ & $\mu_h$ & $\mu_l$\\ \midrule
        3 & $[\frac{1}{3}, \frac{1}{3}, \frac{1}{3}]$ & $[\frac{1}{3}, \frac{1}{3}, \frac{1}{3}]$ & 0.31 & 0.62\\
        4 & $[\frac{1}{4}, \frac{1}{4}, \frac{1}{4}, \frac{1}{4}]$ & $[\frac{1}{4}, \frac{1}{4}, \frac{1}{4}, \frac{1}{4}]$ & 0.34 & 0.51\\
        5 & $[\frac{1}{5}, \frac{1}{5}, \frac{1}{5}, \frac{1}{5}, \frac{1}{5}]$ & $[\frac{1}{5}, \frac{1}{5}, \frac{1}{5}, \frac{1}{5}, \frac{1}{5}]$ & 0.40 & 0.53\\
        6 & $[\frac{1}{6}, \frac{1}{6}, \frac{1}{6}, \frac{1}{6}, \frac{1}{6}, \frac{1}{6}]$ & $[\frac{1}{6}, \frac{1}{6}, \frac{1}{6}, \frac{1}{6}, \frac{1}{6}, \frac{1}{6}]$ & 0.45 & 0.57\\
		\bottomrule
	\end{tabular}
\end{table}
We notice that for a given network load, both $\mu_h$ and $\mu_L$ increase with $M$.
This is due to a reduction in the overloading factor, which is defined as the ratio between the network load and the number of \glspl{RB}.

\subsection{Access Class Baring}
\label{subsec: ACB}
\textcolor{black}{The ACB scheme we adopt bars \glspl{UE} to maintain an overloading factor of 1 that is, the number of \glspl{UE} matches the number of \glspl{RB}. \footnote{The barring rate is chosen to ensure an overloading factor of 1. However, the specific method used to determine this rate is beyond the scope of this work.}
This scheme preserves the ratio of \glspl{H-UE} to \glspl{L-UE} after barring and utilizes uniform \gls{RB} allocation for both \glspl{H-UE} and \glspl{L-UE}.
The results for using \gls{ACB} are shown in \cref{tab: ACB optimization results}.}
\begin{table}
	\footnotesize
	\centering
	\caption{Access class baring results with $n^h=4$, $n^l=5$.}
	\label{tab: ACB optimization results}
	\begin{tabular}{@{}llllll@{}}
		\toprule
		~ & \multicolumn{2}{l}{\textbf{Output Variables}} & \multicolumn{2}{l}{\textbf{Metrics}}\\ \cmidrule(lr){2-3} \cmidrule(lr){4-6}
		$M$ & $\mathbf{p}^h$ & $\mathbf{p}^l$ & $\mu_h$ & $\mu_l$\\ \midrule
        3 & $[\frac{1}{3}, \frac{1}{3}, \frac{1}{3}]$ & $[\frac{1}{3}, \frac{1}{3}, \frac{1}{3}]$ & 0.44 & 0.89\\
        4 & $[\frac{1}{4}, \frac{1}{4}, \frac{1}{4}, \frac{1}{4}]$ & $[\frac{1}{4}, \frac{1}{4}, \frac{1}{4}, \frac{1}{4}]$ & 0.42 & 1.27\\
        5 & $[\frac{1}{5}, \frac{1}{5}, \frac{1}{5}, \frac{1}{5}, \frac{1}{5}]$ & $[\frac{1}{5}, \frac{1}{5}, \frac{1}{5}, \frac{1}{5}, \frac{1}{5}]$ & 0.82 & 1.23\\
        6 & $[\frac{1}{6}, \frac{1}{6}, \frac{1}{6}, \frac{1}{6}, \frac{1}{6}, \frac{1}{6}]$ & $[\frac{1}{6}, \frac{1}{6}, \frac{1}{6}, \frac{1}{6}, \frac{1}{6}, \frac{1}{6}]$ & 0.80 & 1.6\\
		\bottomrule
	\end{tabular}
\end{table}
\textcolor{black}{The results show that \gls{ACB} significantly improves both \gls{RACH} throughputs compared to the uniform \gls{RB} allocation. 
Since \gls{ACB} bars both \glspl{H-UE} and \glspl{L-UE} to maintain an overloading factor of 1, and given that there are initially more \glspl{L-UE} than \glspl{H-UE}, the proportion remains skewed even after barring. 
As a result, \glspl{L-UE} achieve a higher \gls{RACH} throughput than \glspl{H-UE}.}

\subsection{Access Probability Optimization Results}
\label{subsec: RACH optimization results}
First, we present the results of an unconstrained base case to demonstrate the validity of the optimization algorithm in \cref{subsec: problem formulation}. 
Next, we introduce a constraint on the \gls{L-UE} \gls{RACH} throughput and analyze the network's performance.

\subsubsection{Base Case}
\label{subsubsec: base case}
In the unconstrained base case where $\gamma$ is set to zero, we want to ensure that the \glspl{H-UE} receive the maximum share of \glspl{RB}.
Thus, the \glspl{L-UE} should be allocated the minimum number of \glspl{RB}, which is one \gls{RB}, i.e., $p^l_i = 0 \quad \forall i \in \{1, 2, \dots, M-1\} \text{ and } p^l_M = 1$.
The problem can now be viewed as a single priority network with $n^h$ \glspl{H-UE} and $M-1$ \glspl{RB}.
It has been shown that the optimal solution for a single-priority network is to allocate the \glspl{RB} uniformly and to limit the number of \glspl{UE} accessing the network to be equal to $M$, where \gls{ACB} is used to limit the number of \glspl{UE}~\cite{lin_prada_2014}.

\begin{table}[t]
    \footnotesize
	\centering
	\centering \caption{Optimal access probability allocation results with $n^h=4$, $n^l=5$, and $\gamma=0$.}
	\label{tab: unconstrained optimization results}
	\begin{tabular}{@{}llllll@{}}
		\toprule
		~ & \multicolumn{2}{l}{\textbf{Output Variables}} & \multicolumn{3}{l}{\textbf{Metrics}}\\ \cmidrule(lr){2-3} \cmidrule(lr){4-6}
		$M$ & $\mathbf{p}^h$ & $\mathbf{p}^l$ & $\mu_h$ & $\mu_l$ & Time Taken (s)\\ \midrule
		3 & $[\frac{1}{4}, \frac{1}{4}, \frac{1}{2}]$ & $[0, 0, 1]$ & 0.84 & 0 & 4.59\\
		4 & $[\frac{1}{4}, \frac{1}{4}, \frac{1}{4}, \frac{1}{4}]$ & $[0, 0, 0, 1]$ & 1.27 & 0 & 43.85\\
		5 & $[\frac{1}{4}, \frac{1}{4}, \frac{1}{4}, \frac{1}{4}, 0]$ & $[0, 0, 0, 0, 1]$ & 1.68 & 0 & 337.42\\
		6 & $[\frac{1}{5}, \frac{1}{5}, \frac{1}{5}, \frac{1}{5}, \frac{1}{5}, 0]$ & $[0, 0, 0, 0, 0, 1]$ & 2.05 & 0 & 664.51\\ \bottomrule
	\end{tabular}
\end{table}

\textcolor{black}{The results presented in \cref{tab: unconstrained optimization results} show that the \gls{H-UE} \gls{RACH} throughput is significantly higher compared to both the uniform \gls{RB} allocation (\cref{tab: uniform optimization results}) and \gls{ACB}. However, this comes at the cost of the \gls{L-UE} \gls{RACH} throughput being reduced to zero.}
They also confirm that, when $n^h \leq M-1$, the optimal solution is to allocate the $M-1$ \glspl{RB} uniformly, i.e., $p^h_i = \frac{1}{M-1} \quad \forall i \in \{1, 2, \dots, M-1\}$ and reserve the last \gls{RB} only for \glspl{L-UE}, i.e., $p^h_M = 0$.
However, when $n^h > M-1$, the last \gls{RB} is utilized to bar both \glspl{L-UE} and a portion of the \glspl{H-UE} from accessing the network, ensuring that the number of \glspl{UE} accessing the network is equal to $M-1$. In other words, $n^h \sum_{i=1}^{M-1} p^h_i = M-1$.
The last \gls{RB} can be seen as a ``soft'' form of \gls{ACB} to restrict the number of \glspl{UE} accessing the network.

Finally, we observe that as the number of \glspl{RB} increases, the \gls{H-UE} \gls{RACH} throughput increases, which is anticipated since there are more \glspl{RB} to be allocated to the \glspl{H-UE}.
Additionally, the rate of increase of the \gls{H-UE} \gls{RACH} throughput decreases as the number of \glspl{RB} increases due to the law of diminishing returns, i.e., the marginal benefit of adding a \gls{RB} decreases as the number of \glspl{RB} increases.

\subsubsection{Constrained Optimization}
\label{subsubsec: constrained optimization}
In \cref{tab: constrained optimization results}, we present the results for the constrained optimization case where we set the constraint on the \gls{L-UE} \gls{RACH} throughput to 0.4, i.e., $\gamma = 0.4$.
As expected, the \gls{H-UE} \gls{RACH} throughput is lower than the unconstrained case since the \glspl{RB} are now shared between the \glspl{H-UE} and \glspl{L-UE}.
Also, we can see that the \gls{L-UE} \gls{RACH} throughput is equal to 0.4 for all cases, which satisfies the constraint.
\begin{table*}[t]
\footnotesize
    \centering \caption{Optimization results with $n^h=4$, $n^l=5$, and $\gamma=0.4$.}
    \label{tab: constrained optimization results}
    \begin{tabular}{@{}llllll@{}}
        \toprule
        ~ & \multicolumn{2}{l}{\textbf{Output Variables}} & \multicolumn{3}{l}{\textbf{Metrics}}\\ \cmidrule(lr){2-3} \cmidrule(lr){4-6}
        $M$ & $\mathbf{p}^h$ & $\mathbf{p}^l$ & $\mu_h$ & $\mu_l$ & Time Taken (s)\\ \midrule
        3 & [0.006, 0.25, 0.744] & [0.195, 0, 0.805] & 0.43 & 0.4 & 13.06 \\
        4 & [0.25, 0.006, 0.493, 0.25] & [0, 0.196, 0.804, 0] & 0.85 & 0.4 & 75.12 \\
        5 & [0.007, 0.24, 0.251, 0.251, 0.251] & [0.198, 0.802, 0, 0, 0] & 1.28 & 0.4 & 375.25 \\
        6 & [0.01, 0.247, 0.247, 0.247, 0.247, 0] & [0.201, 0, 0, 0, 0, 0.799] & 1.7 & 0.4 & 1281.3 \\
        \bottomrule
    \end{tabular}
\end{table*}

Comparing the uniform \gls{RB} allocation results  (\cref{tab: uniform optimization results}) to the constrained optimization results (\cref{tab: constrained optimization results}), we can see that the \gls{H-UE} \gls{RACH} throughput is higher using the proposed optimization algorithm, while still maintaining the \gls{L-UE} \gls{RACH} throughput above the desired threshold.
\textcolor{black}{Compared to \gls{ACB} (\cref{tab: ACB optimization results}), where the \gls{H-UE} \gls{RACH} throughput ranges from 0.42 to 0.80 and the \gls{L-UE} \gls{RACH} throughput ranges from 0.89 to 1.6, the proposed optimization achieves significantly higher \gls{H-UE} throughput (ranging from 0.43 to 1.7) while keeping \gls{L-UE} throughput stable at 0.4.
This demonstrates that the optimization prioritizes \gls{H-UE} access while maintaining fairness for \gls{L-UE}.}
In both the constrained and unconstrained optimization problems, the time taken to solve the problem increases exponentially with the number of \glspl{RB}, and the time taken to solve the constrained problem is slightly higher than the optimization problem.
Thus, the optimization problem is not scalable for large networks.

\subsection{Multi-Armed Bandits Formulation Results}
\label{subsec: MAB optimization results}
We now present the results of the \gls{MAB}-based proposed method, which are divided into four parts.
Once again, we start with an unconstrained base case, presented in \cref{subsubsec: MAB base case}.
Then, \cref{subsubsec: constrained MAB} introduces the constraint on the \gls{L-UE} \gls{RACH} and analyze the \gls{MAB} performance.
Afterwards, we utilize the \gls{CAS}, presented in \cref{subsubsec: compact AS}, to reduce the \gls{AS} size and analyze the results in \cref{subsubsec: compact AS results}.
Finally, we analyze the impact of varying network load on the \gls{MAB} formulation in \cref{subsubsec: varying network load results}.

As in the previous section, $n^h$ and $n^l$ are set to 4 and 5, respectively, and the number of \glspl{RB} is varied from 3 to 6.
The parameters for the \gls{MAB} formulation are as shown in \cref{tab: MAB parameters}, unless stated otherwise.
The parameters are chosen based on empirical results, where the \gls{MAB} formulation achieves a good balance between accuracy and time taken to solve the optimization problem.

\begin{table}
    \footnotesize
	\centering
	\caption{\Gls{MAB} parameters.}
	\label{tab: MAB parameters}
	\begin{tabular}{@{}ll@{}}
		\toprule
		\textbf{Parameter} & \textbf{Value}\\ \midrule
		$\alpha$ & 0.2\\
		$s$ & 0.1\\
		$B_s$ & 500\\
		$\rho$ & 0\\
		$T$ & 1000\\
		$d$ & 0.2\\
		runs & 15000\\ \bottomrule
	\end{tabular}
\end{table}

\subsubsection{Base Case}
\label{subsubsec: MAB base case}
The results for the unconstrained ($\gamma = 0$) \gls{MAB} formulation are shown in \cref{tab: MAB unconstrained optimization results} for $M=3, \ldots, 6$.
We first note that the $\mu_{l, T}$ is equal to 0 for all cases, which is expected since the problem is unconstrained.
Moreover, comparing the results in \cref{tab: MAB unconstrained optimization results} to the results in \cref{tab: unconstrained optimization results}, we can see that $\mu_{h, T}$ is very close to the optimal solution for the unconstrained problem.
Due to $d$ being set to 0.2, the \gls{MAB} formulation cannot achieve the optimal solution for the unconstrained optimization problem.
However, this is a trade-off between the accuracy of the \gls{MAB} formulation and the size of the \gls{AS}, which directly affects the time taken to solve the optimization problem.
\begin{table}[t]
    \footnotesize
	\centering
	\caption{\Gls{MAB} results with $n^h=4$, $n^l=5$, and $\gamma=0$.}
	\label{tab: MAB unconstrained optimization results}
    \begin{adjustbox}{max width=0.48\textwidth}
	\begin{tabular}{@{}llllll@{}}
		\toprule
		~ & \multicolumn{2}{l}{\textbf{Output Variables}} & \multicolumn{3}{l}{\textbf{Metrics}}\\ \cmidrule(lr){2-3} \cmidrule(lr){4-6}
		$M$ & $\mathbf{p}^h$ & $\mathbf{p}^l$ & $\mu_{h, T}$ & $\mu_{l, T}$ & Time Taken (s)\\ \midrule
		3 & $[0.2, 0.2, 0.6]$ & $[0, 0, 1]$ & 0.82
		& 0 & 38.10 \\
		4 & $[0.2, 0.2, 0.2, 0.4]$ & $[0, 0, 0, 1]$ & 1.23 & 0 & 62.79\\
		5 & $[0.2, 0.2, 0.2, 0.4, 0]$ & $[0, 0, 0, 0, 1]$ & 1.57 & 0 & 96.82\\
		6 & $[0.2, 0.2, 0.2, 0.2, 0.2, 0]$ & $[0, 0, 0, 0, 0, 1]$ & 2.05 & 0 & 196.99\\ \bottomrule  
	\end{tabular}
    \end{adjustbox}
\end{table}

\begin{figure}[t]
	\centering
	\includegraphics[width=0.48\textwidth]{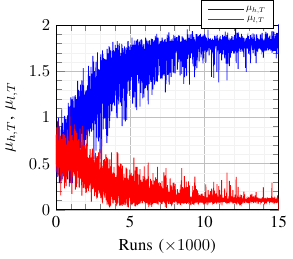}
	\centering 
    \caption{$\mu_{h, T}$ and $\mu_{l, T}$ versus runs for $\gamma=0$, $M=6$.}
	\label{fig: MAB unconst}
\end{figure}

\cref{fig: MAB unconst} shows $\mu_{h, T}$ and $\mu_{l, T}$ versus the number of runs for $M=6$.
We can see that the \gls{MAB} formulation converges to the optimal solution for the unconstrained problem, with $\mu_{h, T}$ reaching 2.05 and $\mu_{l, T}$ reaching 0.
Furthermore, the time taken to solve the optimization problem is not directly dependent on the number of \glspl{RB} but rather on the \gls{MAB} parameters, which are not changed for different values of $M$.
Comparing the time taken to solve the constrained optimization problem using the \gls{MAB} formulation to the time taken to solve the constrained optimization problem using the optimal access probability allocation, we can see that for a very small network, the \gls{MAB} formulation is slower than the optimal access probability allocation.
However, as the network size increases, the \gls{MAB} formulation becomes significantly faster than the optimal access probability allocation.

\subsubsection{Constrained Multi-Armed Bandits}
\label{subsubsec: constrained MAB}
We now introduce the constraint on the \gls{L-UE} \gls{RACH} throughput, $\gamma=0.4$, and analyze the \gls{MAB} formulation's performance.
The results are shown in \cref{tab: MAB constrained optimization results}.
We can see that the \gls{MAB} formulation satisfies the constraint on the \gls{L-UE} \gls{RACH} throughput, i.e., $\mu_{l, T}= 0.41$ for all cases.
Looking at $\mathbf{p}^l$, we can see that the non-uniform access probabilities are very close to the optimal solution for the constrained problem with slight deviations due to the discretization of the \gls{AS}.
Moreover, the values of $\mu_{h, T}$ are within 5\% of the optimal solution for the constrained problem. 
\begin{table}[t]
    \footnotesize
	\centering
    \caption{\Gls{MAB} results with $n^h=4$, $n^l=5$, and $\gamma=0.4$.}
	\label{tab: MAB constrained optimization results}
    \begin{adjustbox}{max width=0.48\textwidth}
	\begin{tabular}{@{}llllll@{}}
		\toprule
		~ & \multicolumn{2}{l}{\textbf{Output Variables}} & \multicolumn{3}{l}{\textbf{Metrics}}\\ \cmidrule(lr){2-3} \cmidrule(lr){4-6}
		$M$ & $\mathbf{p}^h$ & $\mathbf{p}^l$ & $\mu_{h, T}$ & $\mu_{l, T}$ & Time Taken (s)\\ \midrule
		3 & $[0, 0.2, 0.8]$ & $[0.2, 0, 0.8]$ & 0.41
		& 0.41 & 674.45 \\
		4 & $[0, 0.2, 0.6, 0.2]$ & $[0.2, 0, 0, 0.8]$ & 0.82 & 0.41 & 670.74\\
		5 & $[0, 0.2, 0.2, 0.2, 0.4]$ & $[0.2, 0, 0, 0, 0.8]$ & 1.23 & 0.41 & 672.28\\
		6 & $[0, 0.2, 0.2, 0.2, 0.2]$ & $[0.2, 0, 0, 0, 0, 0.8]$ & 1.64 & 0.41 & 683.81\\ \bottomrule  
	\end{tabular}
    \end{adjustbox}
\end{table}

\begin{figure*}[t] %
    \centering
    \subfloat[]{
		\includegraphics[width=0.45\textwidth]{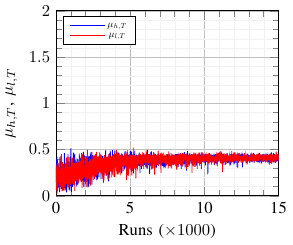}
        \label{fig: MAB const M=3}
        } \hspace{0.25cm}
    \subfloat[]{
		\includegraphics[width=0.45\textwidth]{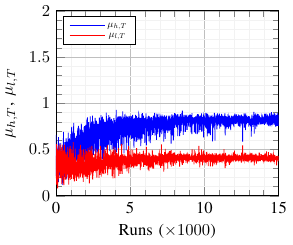}
        \label{fig: MAB const M=4}
        } \hspace{0.25cm}

    \subfloat[]{
		\includegraphics[width=0.45\textwidth]{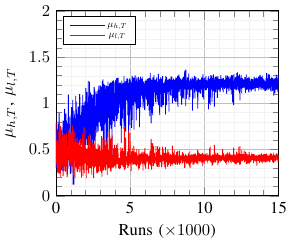}
        \label{fig: MAB const M=5}
        } \hspace{0.25cm}
    \subfloat[]{
		\includegraphics[width=0.45\textwidth]{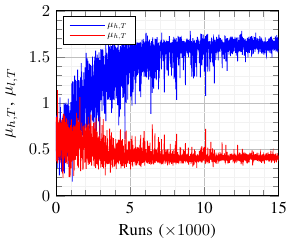}
        \label{fig: MAB const M=6}
        }
    \centering \caption{$\mu_{h, T}$ and $\mu_{l, T}$ versus runs for $\gamma = 0.4$ and (\ref{fig: MAB const M=3}) $M=3$, (\ref{fig: MAB const M=4}) $M=4$, (\ref{fig: MAB const M=5}) $M=5$, (\ref{fig: MAB const M=6}) $M=6$.}
    \label{fig: MAB const}
\end{figure*}

\cref{fig: MAB const} shows $\mu_{h, T}$ and $\mu_{l, T}$ versus the number of runs for $M=3, \ldots, 6$.
We can see that as $M$ increases, the difference between $\mu_{l, T}$ and $\mu_{l, T}$ increases due to the availability of more \glspl{RB}.
The time it takes to solve the constrained optimization problem follows the same trend as the time taken to solve the unconstrained optimization problem.

\subsubsection{Compact Action Space Multi-Armed Bandits}
\label{subsubsec: compact AS results}
We now present the results of the \gls{CAS} \gls{MAB} formulation.
The simulation parameters are shown in \cref{tab: MAB parameters with optimal AS}, and are chosen based on empirical results, where the \gls{MAB} formulation achieves a good balance between accuracy and time taken to solve the optimization problem.
For the \gls{CAS}, we sweep both $n^h$ and $n^l$ from 0 to 10 and save the optimal access probabilities for each case in a lookup table, resulting in an \gls{AS} of size $11 \times 11$.
This is repeated for all values of $M$ from 3 to 6.
Note that there is no need to set $d$ for the \gls{CAS} \gls{MAB} formulation since the \gls{AS} is already discrete in nature.
\begin{table}
    \footnotesize
	\centering
	\centering \caption{\Gls{MAB} parameters with the \gls{CAS}.}
	\label{tab: MAB parameters with optimal AS}
	\begin{tabular}{@{}ll@{}}
		\toprule
		\textbf{Parameter} & \textbf{Value}\\ \midrule
		$\alpha$ & 0.1\\
		$s$ & 0.1\\
		$B_s$ & 200\\
		$\rho$ & 0.1\\
		$T$ & 100\\
		runs & 2000\\ \bottomrule
	\end{tabular}
\end{table}

The access probabilities obtained by this \gls{MAB} formulation are the optimal ones, previously presented in \cref{tab: constrained optimization results}.
However, the time taken to solve the problem, as shown in \cref{tab: constrained optimization results with optimal AS}, is significantly lower, especially for larger values of $M$ when compared to the optimal access probability allocation and the previous \gls{MAB} formulation.
This is attributed to the reduced size search space, which directly affects the time taken to solve the optimization problem.
\begin{table}
	\footnotesize
	\centering
	\centering \caption{Time taken for \gls{MAB} using the \gls{CAS}.}
	\label{tab: constrained optimization results with optimal AS}
	\begin{tabular}{@{}ll@{}}
		\toprule
		\textbf{Parameter} & \textbf{Metric}\\ \cmidrule(lr){1-2}
		$M$ & Time Taken (s) \\ \midrule
		3 & 54.88 \\
		4 & 58.80 \\
		5 & 65.24 \\
		6 & 78.51 \\ \bottomrule
	\end{tabular}
\end{table}

\begin{figure*}[t] %
    \centering
    \subfloat[]{
		\includegraphics[width=0.45\textwidth]{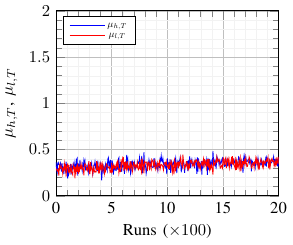}
        \label{fig: MAB optimal AS const M=3}
        } \hspace{0.25cm}
    \subfloat[]{
		\includegraphics[width=0.45\textwidth]{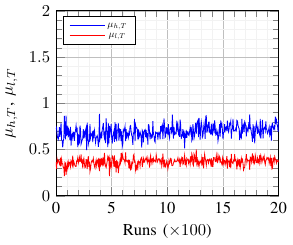}
        \label{fig: MAB optimal AS const M=4}
        } \hspace{0.25cm}

    \subfloat[]{
		\includegraphics[width=0.45\textwidth]{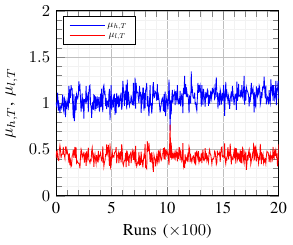}
        \label{fig: MAB optimal AS const M=5}
        } \hspace{0.25cm}
    \subfloat[]{
		\includegraphics[width=0.45\textwidth]{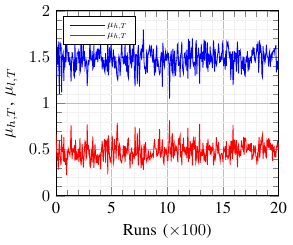}
        \label{fig: MAB optimal AS const M=6}
        }
    \centering \caption{$\mu_{h, T}$ and $\mu_{l, T}$ versus runs for $\gamma = 0.4$ and (\ref{fig: MAB const M=3}) $M=3$, (\ref{fig: MAB const M=4}) $M=4$, (\ref{fig: MAB const M=5}) $M=5$, (\ref{fig: MAB const M=6}) $M=6$ using the \gls{CAS}.}
    \label{fig: MAB const optimal AS}
\end{figure*}

\cref{fig: MAB const optimal AS} shows $\mu_{h, T}$ and $\mu_{l, T}$ versus the number of runs for $M=3$ to $6$.
We can see that the convergence of the \gls{MAB} formulation with the \gls{CAS} is almost instant and much faster than the previous \gls{MAB} formulation.

\subsubsection{Varying Network Load}
\label{subsubsec: varying network load results}
\textcolor{black}{We now analyze the performance of the \gls{MAB} formulation for varying network loads for both the discretized \gls{AS} and the \gls{CAS}.}

\textcolor{black}{For the discretized \gls{AS}, the number of runs is set to 30,000, with the network load configured as $n^h=2$, $n^l=1$ for the first 15,000 runs and $n^h=4$, $n^l=5$ for the next 15,000.
Similarly, for the \gls{CAS}, 4,000 runs are performed, with the same load settings applied over the first and second 2,000 runs, respectively.
All other parameters follow those in \cref{tab: MAB parameters} for discretized \gls{AS} and \cref{tab: MAB parameters with optimal AS} for \gls{CAS}.
This setup simulates a dynamic network where the number of \glspl{H-UE} and \glspl{L-UE} varies over time.}

\textcolor{black}{For these simulation parameters, the maximum achievable \gls{H-UE} \gls{RACH} throughput while ensuring that the \gls{L-UE} \gls{RACH} throughput remains above 0.4 is 1.4 for \gls{H-UE} when $n^h=2$, $n^l=1$, with an \gls{L-UE} throughput of 1.0.
The high \gls{L-UE} value in this case is due to the overloading factor being less than 1, meaning there are more \glspl{RB} than \glspl{UE}. 
For $n^h=4$, $n^l=5$, the maximum \gls{H-UE} throughput is 1.2282, with the \gls{L-UE} throughput constrained to 0.4.}

\begin{figure}[t]
	\centering
    \subfloat[]{
		\includegraphics[width=0.45\textwidth]{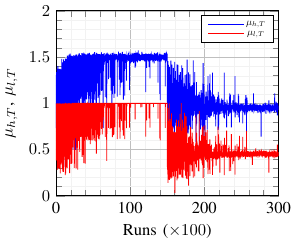}
	\label{fig: MAB varying network load DAS}
    } \hspace{0.25cm}
    \subfloat[]{
		\includegraphics[width=0.45\textwidth]{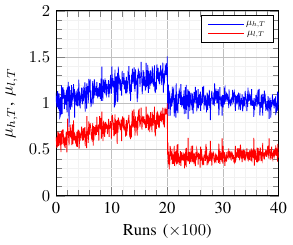}
	\label{fig: MAB varying network load CAS}
    }
	\centering \caption{$\mu_{h, T}$ and $\mu_{l, T}$ versus runs for $M=5$ under varying network load using (\ref{fig: MAB varying network load DAS}) the discretized \gls{AS} and (\ref{fig: MAB varying network load CAS}) the \gls{CAS}.}
	\label{fig: MAB varying network load}
\end{figure}

\textcolor{black}{The result for $M=5$ is shown in \cref{fig: MAB varying network load}.
Observing \cref{fig: MAB varying network load}, we clearly see the sudden change in the network load at run 15,000 and run 2,000 for the discretized \gls{AS} and the \gls{CAS}, respectively.
Futhermore, notice that both \gls{MAB} formulations are able to adapt to the varying network load, with $\mu_{h, T}$ converging to the maximum possible value while $\mu_{l, T}$ remains above the required threshold. This demonstrates the effectiveness of both \gls{MAB} formulations in dynamically adjusting to changes in network conditions while ensuring that the \gls{L-UE} throughput constraint is met.}
\textcolor{black}{We finally observe that the \gls{CAS} converges faster than the discretized action space, highlighting its efficiency in responding to network dynamics.
However, as previously stated, this advantage comes with the tradeoff of requiring two additional hyperparameters to be carefully selected, that are based on rough network load estimates.}

\subsection{Network Load Estimation Using Multi-Armed Bandits}
\label{subsec: MAB estimation}
\begin{figure}[ht]
	\centering
	\includegraphics[width=0.45\textwidth]{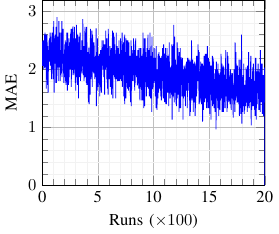}
	\centering \caption{\gls{MAE} versus runs for $\gamma = 0.4$ and $M=6$.}
	\label{fig: MAB estimation}
\end{figure}

We now present the results of the \gls{MAB} formulation for estimating the network load using the \gls{CAS} and referencing the lookup table.
The parameters for this formulation are as in \cref{tab: MAB parameters with optimal AS} and $\gamma$ is set to 0.4.
\cref{fig: MAB estimation} shows the \gls{MAE} versus runs for $M=6$.
We observe that the \gls{MAE} is relatively small when compared to our work in~\cite{elmeligy2024load} and there is a general trend of decreasing \gls{MAE} as the number of runs increases.

\section{Conclusion}
\label{sec: conclusion}
In this paper, we proposed a novel two-priority \gls{RACH} scheme that utilizes non-uniform access probabilities to improve the \gls{RACH} throughput of \glspl{H-UE} while maintaining a minimum one for \glspl{L-UE}.
We formulated the problem as an optimization problem to be solved so that the optimal access probabilities using \gls{SQP} can be found.
Furthermore, a \gls{MAB} formulation that discretizes the \gls{AS} is proposed to solve the optimization problem in a more time-efficient and scalable manner.
A \gls{CE} based algorithm along with a size reduction technique are proposed to tackle the large \gls{AS} size.
Moreover, a \gls{CAS} \gls{MAB} formulation is proposed to reduce the search space and solve the optimization problem more efficiently, with a byproduct of estimating the network load.

The results show that our formulation surpases the traditional uniform access probability scheme in terms of \gls{RACH} throughput for the \glspl{H-UE}, while maintaining a minimum \gls{RACH} throughput for the \glspl{L-UE}.
Moreover, the \gls{MAB} formulations are within 5\% of the optimal solution for the constrained problem while being significantly faster to compute than solving the optimal access probability problem.

An interesting direction for future research is to consider the impact of the estimation process, where the \gls{BS} does not initially know the number of \glspl{H-UE} and \glspl{L-UE} in the network. The \gls{BS} first needs to estimate these numbers before optimizing the access probabilities. Future studies could explore the effects of estimation errors on the optimization problem, conduct a sensitivity analysis to understand how robust the solutions are to these errors, and investigate how different \glspl{RB} access probabilities influence the accuracy of the estimations.
\textcolor{black}{Finally, we could explore varying \gls{UE} activation distributions, such as using uniform or beta distributions for activation times~\cite{3GPP}. This approach would allow for a deeper understanding of their effects on network performance and resource allocation. By considering different distributions, future studies could simulate more realistic activation scenarios, offering insights into how such variations influence system efficiency, reliability, and overall robustness.}

\appendices

\section{Action Space Discretization Algorithms}
\label{app: AS discretization algorithms}
In this appendix, we describe the generation of $\mathcal{AS}$, and how the majority of redundant actions are eliminated by adding a circular shift check.
The algorithm, presented in \cref{alg: discretization with circular shift check}, takes as input the number of \glspl{RB} ($M$), the discretization step ($d$), the intermediate vector ($\mathbf{s}$), and an empty list $\mathcal{E}$, which is later used to store existing combinations.
The algorithm returns the set $\mathcal{AS}$, where each element consists of two vectors, $\mathbf{p}^h$ and $\mathbf{p}^l$.

\begin{algorithm}
	\centering \caption{Discretization function with circular shift check}
	\label{alg: discretization with circular shift check}
	\footnotesize
	\begin{algorithmic}[1]
		\Function{DiscretizeAsCs}{$M, r, \mathbf{s}$, $\mathcal{E}$}
			\State $\mathcal{AS} \gets \text{empty list}$
			\State $\mathbf{p}^h_p=\mathbf{s}[0]$, $\mathbf{p}^l_p=\mathbf{s}[1]$
			\If{$\sum \mathbf{p}^h_p = 1$ \textbf{and} $\sum \mathbf{p}^l_p = 1$ \textbf{and} $M=0$ \textbf{and not} $\texttt{IsCircularShift}(\mathbf{s}, \mathcal{E})$} \label{alg: discretization base case reduced}
				\State $\mathcal{E}.\text{append}(\mathbf{s}.\text{copy}())$ \label{alg: discretization base case 1}
				\State $\mathcal{AS}.\text{append}(\mathbf{s}.\text{copy}())$ \label{alg: discretization base case 2}
				\State \Return $\mathcal{AS}$
			\EndIf
			\If{$\sum \mathbf{p}^h_p > 1$ \textbf{or} $\sum \mathbf{p}^l_p > 1$ \textbf{or} $M \leq 0$}
				\State \Return $\mathcal{AS}$ \label{alg: discretization pruning}
			\EndIf
			\For{$i$ in $\text{.arange}(0, 1 + r, r)$} \label{alg: discretization recursive exploration start}
				\For{$j$ in $\text{.arange}(0, 1 + r, r)$}
					\State $\mathbf{s}[0].\text{append}(i)$
					\State $\mathbf{s}[1].\text{append}(j)$ \label{alg: discretization recursive exploration end}
					\State $\mathcal{N} \gets \texttt{DiscretizeAsCs}(M - 1, r, \mathbf{s}, \mathcal{E})$ \label{alg: discretization recursive call}
					\State $\mathcal{AS}.\text{extend}(\mathcal{N})$ \label{alg: discretization append}
				\EndFor
			\EndFor
			\State \Return $\mathcal{AS}$
		\EndFunction
	\end{algorithmic}
\end{algorithm}

\cref{alg: discretization with circular shift check} utilitez backtracking hence it has three distinct phases.
The handling of the base case of \cref{alg: discretization with circular shift check} is met if: 1) The sum of the elements in $\mathbf{p}^h_p$ and $\mathbf{p}^l_p$ equals one. 2) The remaining \glspl{RB} are zero. 3) The current solution is not a circular shift of any of the existing solutions saved in $\mathcal{E}$ (line~\ref{alg: discretization base case reduced}).
The first condition ensures that the probability vectors are valid.
The second ensures that the length of $\mathbf{p}^h_p$ and $\mathbf{p}^l_p$ is the same as the number of \glspl{RB}, that is, every \gls{RB} has an access probability associated with it.
The third condition is to ensure that the generated solution is not a circular shift of any of the existing solutions saved in $\mathcal{E}$ by calling a circular shift check, explained later.

In the event that the base case conditions are satisfied, $\mathbf{s}$ is appended to the list of existing solutions and the output set, which constitutes a valid solution branch (lines~\ref{alg: discretization base case 1} and \ref{alg: discretization base case 2}).
Next, the algorithm prunes infeasible solution branches by checking whether the sum of the elements in $\mathbf{p}^h_p$ or $\mathbf{p}^l_p$ is greater than 1 or whether $M$ is less than or equal to 0 (line~\ref{alg: discretization pruning}).
If any of the above conditions are met, the algorithm does not explore the current branch further and returns the current set of solutions.
Finally, if the base case and pruning conditions are not satisfied, the current branch is further explored by appending the next possible values of $\mathbf{p}^h_p$ and $\mathbf{p}^l_p$, depending on $d$, to $\mathbf{s}$ (lines~\ref{alg: discretization recursive exploration start} to \ref{alg: discretization recursive exploration end}).
The algorithm then makes a recursive call to itself with the updated values of $M$, $\mathbf{s}$, and $\mathcal{E}$ and
appends the output of the recursive call, $\mathcal{N}$, to the output set (lines~\ref{alg: discretization recursive call} and \ref{alg: discretization append}).
The algorithm repeats this process for all combinations of $\mathbf{p}^h$ and $\mathbf{p}^l$ until the set $\mathcal{AS}$ is returned that contains all possible actions without circular shift redundancy.

\cref{alg: is circular shift}  describes how the circular shift check is performed; it takes as input the current combination $\mathbf{s}$ and the list of possible existing combinations $\mathcal{E}$.
Two distinct phases are involved in the circular shift check: circular shift generation and circular shift comparison.
In the circular shift generation phase, a set of circular shifts, $\mathcal{CS}$, of the current combination $\mathbf{s}$ is generated (lines~\ref{alg: generate circular shifts start} to \ref{alg: generate circular shifts end}).
In the circular shift comparison phase, the algorithm compares each circular shift in $\mathcal{CS}$, denoted by $\mathbf{cs}$, to each action in $\mathcal{E}$, denoted by $\mathbf{e}$, to determine whether the current combination is a circular shift of an existing combination.

\begin{algorithm}
	\centering \caption{Check for Circular Shifts}
	\label{alg: is circular shift}
	\footnotesize
	\begin{algorithmic}[1]
		\Function{IsCircularShift}{$\mathbf{s}, \mathcal{E}$}
			\State $\mathcal{CS} \gets \text{empty list}$
			\For{$\mathcal{CS}$ in $\text{range}(|\mathbf{s}[0]|)$} \label{alg: generate circular shifts start}
				\State $\mathbf{p}^h \gets \text{roll}(|\mathbf{s}[0]|, \mathcal{CS})$
				\State $\mathbf{p}^l \gets \text{roll}(|\mathbf{s}[1]|, \mathcal{CS})$
				\State $\mathcal{CS}\text{.append}([P, X])$
			\EndFor \label{alg: generate circular shifts end}
			\For{$\mathbf{cs}$ in $\mathcal{CS}$} \label{alg: compare circular shifts start}
				\For{$\mathbf{e}$ in $\mathcal{E}$}
					\If{$\mathbf{e} == \mathbf{cs}$}
						\State \Return \texttt{True}
					\EndIf
				\EndFor
			\EndFor
			\State \Return \texttt{False} \label{alg: compare circular shifts end}
		\EndFunction
	\end{algorithmic}
\end{algorithm}

\section{Random Access Channel Simulation}
\label{app: RACH simulation}

\begin{algorithm}[t]
	\centering \caption{\gls{RACH} simulation}
	\label{alg: RACH simulation}
	\footnotesize
	\begin{algorithmic}[1]
		\Function{SimRACH}{$n^h$, $n_l$, $\bm{p}^h$, $\bm{p}^l$, $T$, $M$}
		\State $y \gets \text{empty dictionary}$, $RB \gets \text{range}(M)$ \label{alg: RACH init}		
		\State $C_{n^h} \gets$ Randomly select from $RB$ with size ($n^h$, $T$) and probability~$P^h$ \label{alg: RACH Cnh}
		\State $C_{n^l} \gets$ Randomly select from $RB$ with size ($n^l$, $T$) and probability~$P^l$ \label{alg: RACH Cnl}
		\State $C \gets \text{append} (C_{n^h}, C_{n^l})$ \label{alg: RACH C}

		\State $j \gets 0$ \label{alg: RACH loop 1 counter}
		\For{$c \text{ in } C$} \label{alg: RACH loop 1 start}
			\State $F \gets \text{zeros}(M)$ \label{alg: RACH F init}
			\State $u, f \gets \text{unique}(c)$ \label{alg: RACH unique}
			\State $F[u] \gets f$ \label{alg: RACH F update}
			\State $y[j] \gets \text{dict}(\text{zip}(RBs, F))$ \label{alg: RACH d update}
			\State $y[j] \gets RB: 'x' \text{ if } f > 1 \text{ else } f \text{ for } RB, f \text{ in items.}y[j]$ \label{alg: RACH d update x}
			\State $y[j] \gets RB: 'o' \text{ if } f == 0 \text{ else } f \text{ for } RB, f \text{ in items.}y[j]$ \label{alg: RACH d update o}
			
			\If{$\text{any (F==1)}$} \label{alg: RACH if F==1}
				\State $I \gets \text{where }(F==1)$ \label{alg: RACH I}
				\For{$i \text{ in } I$} \label{alg: RACH loop 2 start}
					\State $UE_i \gets \text{where }(j==i)$ \label{alg: RACH UEi}
					\If{$UE_i < n^h$} \label{alg: RACH if UEi}
						\State $y[j][i] \gets 'h'$ \label{alg: RACH d update h}
					\Else
						\State $y[j][i] \gets 'l'$ \label{alg: RACH d update l}		
					\EndIf
				\EndFor
			\EndIf
			\State $j \gets j + 1$ \label{alg: RACH loop 1 counter update}
		\EndFor
		\State \Return $y$ \label{alg: RACH return}
		\EndFunction
	\end{algorithmic}
\end{algorithm}

This appendix describes the \gls{RACH} simulation setup that is used to obtain $\mu_{h, T}$ and $\mu_{l, T}$ in \cref{alg: MAB formulation}, which is shown in \cref{alg: RACH simulation}.
For this section, we simplify the events presented in \cref{sec: system model} to $h$ for a successful \gls{H-UE} access, $l$ for a successful \gls{L-UE} access, $o$ for an unsuccessful access, and $x$ for a collision (dropping the subscript $i$ for simplicity).
We note that this simplification is for illustrative purposes and does not affect the simulation results.
The inputs to the algorithm are the number of \glspl{RB} ($M$), \glspl{H-UE} ($n^h$), \glspl{L-UE} ($n^l$), trials ($T$), and the \gls{RB} access probabilities for a \gls{H-UE} ($\bm{p}^h$) and an \gls{L-UE} ($\bm{p}^l$).
The algorithm initializes an empty dictionary, $y$, which is used to store the output, i.e., the $T$ access patterns.
An array, $RBs$ from 1 to $M$ is also initialized to represent the \glspl{RB} (line~\ref{alg: RACH init}).
The access patterns for the \glspl{H-UE} and \glspl{L-UE} are generated by randomly selecting $n^h$ and $n^l$ \glspl{RB} from $M$ with probabilities $\bm{p}^h$ and $\bm{p}^l$, which are stored in matrices $C_{n^h}$ and $C_{n^l}$, respectively (lines~\ref{alg: RACH Cnh} and \ref{alg: RACH Cnl}).
Matrix $C$ is created by appending $C_{n^h}$ and $C_{n^l}$ (line~\ref{alg: RACH C}).

Afterwards, the algorithm initializes a counter, $j$, representing the current access pattern, and iterates over the columns of $C$, denoted by $c$ (lines~\ref{alg: RACH loop 1 counter} and \ref{alg: RACH loop 1 start}).
For each column, $c$, the algorithm initializes a frequency vector, $F$, finds the unique elements, $u$, and how many times they occur, $f$ (lines~\ref{alg: RACH F init} and \ref{alg: RACH unique}).
Next, $F$ is updated with $f$ at the index of $u$ (line~\ref{alg: RACH F update}).
The dictionary $y$ is formulated by having $j$ as the key and values that are dictionaries with keys as the $RBs$ and values as the frequency of the \glspl{RB} in the access pattern (line~\ref{alg: RACH d update}).
For each \gls{RB} in access pattern, $j$, the algorithm checks if more than one \gls{UE} is accessing the \gls{RB} and updates $y$ with 'x' if true and 'o' if no \glspl{UE} are accessing the \gls{RB} (lines~\ref{alg: RACH d update x} and \ref{alg: RACH d update o}).
If any \gls{RB} has only one \gls{UE} accessing it, the algorithm checks if the \gls{UE} is a \gls{H-UE} or a \gls{L-UE} and updates $y$ accordingly with either 'h' or 'l', respectively (lines~\ref{alg: RACH if F==1} and \ref{alg: RACH d update l}).
Finally, the algorithm increments $j$, and repeats the process for the next access pattern until all $T$ access patterns are generated, which are returned as $d$ (lines~\ref{alg: RACH loop 1 counter update} and \ref{alg: RACH return}).

\begin{singlespace}
\bibliographystyle{jabbrv_IEEEtran}
\bibliography{References/references_all.bib}
\end{singlespace}

\begin{IEEEbiography}[{\includegraphics[width=1in,height=1.25in,clip,keepaspectratio]{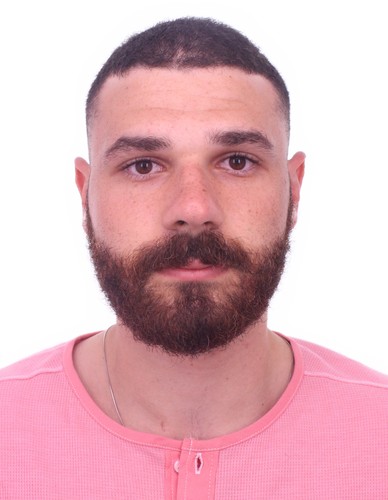}}]{Ahmed O. Elmeligy}
	Ahmed O. Elmeligy received the B.Sc. and M.Sc. degrees in Electrical Engineering from the American University of Sharjah, Sharjah, United Arab Emirates, in 2020 and 2022, respectively. 
	From 2021 to 2022, he was a Research and Development Engineer in robotics and automation in Abu Dhabi, United Arab Emirates.
	He is currently pursuing a Ph.D. in electrical engineering at McGill University, Montreal, QC, Canada. 
	His research interests include wireless communications, optimization, and reinforcement learning.
\end{IEEEbiography}

\begin{IEEEbiographynophoto}{Ioannis Psaromiligkos}
	Ioannis Psaromiligkos received the Diploma degree in computer engineering and science from the University of Patras, Patras, Greece, in 1995, and the M.S. and Ph.D. degrees in electrical engineering from the State University of New York, Buffalo, NY, USA, in 1997 and 2001, respectively. Since 2001, he has been with the Department of Electrical and Computer Engineering, McGill University, Montreal, QC, Canada, where he is currently an Associate Professor. His research interests are in the areas of statistical detection and estimation, adaptive signal processing, machine learning, and wireless communications. He has served as an Associate Editor for IEEE Communications Letters and IEEE Signal Processing Letters. He is currently an Associate Editor of the EURASIP Journal on Advances in Signal Processing.
\end{IEEEbiographynophoto}

\begin{IEEEbiographynophoto}{Au Minh}
	Au Minh is a Researcher Scientist with Hydro-Quebec Research Institute (IREQ), Varennes, QC, Canada. His research interests include the digital transformation of power substations, partial discharge phenomena, information theory, and cyber security for smart grids.
\end{IEEEbiographynophoto}

\end{document}